# Towards a Zero-Trust Supply-Chain Assurance Rubric for O-RAN RIC Applications

Chun Yin Chiu

**Abstract.** Open RAN enables third-party xApps and rApps to be onboarded and updated at operational cadence. This application model increases the importance of software supply-chain assurance across build, signing, publication, onboarding, runtime, and update stages. This paper proposes an app-centric assurance framework for O-RAN RIC applications. The framework makes three contributions: (1) a lifecycle threat model for RIC applications and their trust boundaries; (2) a WG11-aligned threat-control-evidence mapping that links lifecycle threats to O-RAN security baselines and complementary software supply-chain evidence; and (3) an operator-facing assurance profile that combines SSDF practices, SBOM transparency, and SLSA-style provenance into incremental onboarding levels. The evaluation is analytical rather than empirical: three case-study walkthroughs and a minimal checking workflow illustrate how the framework can support explicit evidence-based onboarding decisions. Empirical deployment measurements and inter-reviewer consistency studies are left for future work.

## 1 Introduction

The O-RAN architecture disaggregates the radio access network and promotes open interfaces, cloud-native deployment, and multi-vendor interoperability [1]. A key enabler is the RAN Intelligent Controller (RIC), which hosts third-party applications: xApps in the Near-Real-Time RIC and rApps in the Service Management and Orchestration (SMO) / Non-Real-Time RIC domains [1]. This application model changes the threat landscape. Instead of deploying monolithic vendor network functions on long release cycles, operators may ingest and update app artifacts from a diverse ecosystem, often as container images, packages, models, or policy bundles.

This paper focuses on a practical problem that arises at the boundary between O-RAN security conformance and software supply-chain assurance. WG11 provides security requirements, protocol profiles, threat modeling, and security test specifications for O-RAN components and interfaces [2]-[5]. In parallel, the broader software ecosystem has developed supply-chain integrity practices such as the Secure Software Development Framework (SSDF), software bills of materials (SBOMs), and SLSA-style provenance [8]-[10]. To the best of our knowledge, operators still lack a publicly documented, reusable, app-centric method for connecting these evidence families to RIC app onboarding and update decisions.

The goal is not to replace WG11 conformance practice. Rather, the paper treats WG11 interface and platform security as the baseline and layers complementary app supply-chain evidence on top. The intended outcome is an operator-facing rubric that makes onboarding decisions auditable: each lifecycle risk is mapped to controls, and each control is backed by evidence that can be checked before deployment or monitored at runtime.

The paper makes the following contributions:

1. An app-centric threat model and taxonomy organized around the xApp/rApp lifecycle: build, sign, publish, onboard, runtime, and update/rollback.
2. A zero-trust threat-control-evidence framework aligned with WG11 baselines and complemented by SBOM, signing, and provenance evidence.
3. A RIC App Supply-Chain Assurance Profile with four operator-facing levels, designed for admission policy, procurement guidance, and post-incident audit.

## 2 Background and Scope

In O-RAN, the Near-RT RIC executes control loops on the order of milliseconds to seconds and hosts xApps, while the SMO/Non-RT RIC hosts rApps for non-real-time optimization, orchestration, policy, and AI/ML workflows [1]. The Near-RT RIC communicates with E2 nodes, such as O-CU and O-DU, via the E2 interface and exchanges policies and enrichment information with the Non-RT RIC via A1 [1]. Operationally, xApps and rApps are onboarded into a RIC platform, granted identities and permissions, and then invoke platform APIs to observe telemetry or influence control behavior.

WG11 security deliverables provide the baseline for many interface and platform controls. The core document set used in this paper includes Security Requirements and Controls Specifications [2], Security Protocols Specifications [3], Security Test Specifications [4], and Security Threat Modeling and Risk Assessment [5]. These documents are complementary: requirements define what must be implemented, protocol specifications define how secure communication and authentication are profiled for applicable components, test specifications define validation procedures, and threat modeling provides risk context.

This work is deliberately scoped to the supply chain and operational lifecycle of RIC apps and the O-RAN components directly involved in onboarding, identity, authorization, and runtime isolation for those apps. It does not attempt to exhaustively cover all O-RAN interfaces, all 3GPP security requirements, or all operational security processes. Instead, it asks what additional evidence is useful before a third-party app is allowed to run in an operator-controlled RIC environment.

## 3 System and Adversary Model

We model a representative O-RAN deployment consisting of an SMO/Non-RT RIC domain that manages rApps and policies, a Near-RT RIC domain that hosts xApps and executes near-real-time control loops, and an O-Cloud infrastructure layer that provides compute, storage, network, and platform services [1]. The supply chain spans software producers and their CI systems, artifact signing and publication to a registry, operator onboarding workflows, and runtime enforcement and monitoring.

### 3.1 Assets

Key assets include: app artifacts such as container images, packages, policy logic, and ML models; cryptographic material such as signing keys, certificates, and trust anchors; policy artifacts such as A1 policies, authorization policies, and admission policies; RIC platform services such as app managers, API gateways, and message routers; telemetry and control data such as KPMs, logs, configuration, and intents; and the underlying O-Cloud runtime that enforces isolation.

### 3.2 Adversary classes

Consistent with zero-trust assumptions and WG11 risk-assessment practice, we consider external supply-chain attackers targeting repositories, CI runners, artifact registries, or update channels; compromised dependencies and build tooling, including dependency-confusion attacks; malicious or coerced developers with legitimate credentials; compromised operator tooling; and insiders who misuse privileges or tamper with policies at runtime [5], [6]. The model assumes that network reachability or possession of a developer credential is not sufficient to establish trust.

### 3.3 Trust boundaries

Two boundaries are central. First, artifacts cross the producer-to-operator distribution boundary, where software moves from a supplier-controlled environment into an operator-controlled onboarding pipeline. Second, artifacts cross the onboarding-to-runtime boundary, where app identity is bound to permissions and enforcement points. Each boundary requires explicit verification rather than implicit trust.

*Figure 1. RIC app lifecycle trust boundaries: producer build environment -> signed release -> registry -> operator onboarding -> runtime enforcement -> update/rollback gate.*

## 4 Lifecycle Attack-Surface Taxonomy for xApps/rApps

The attack surface is organized by lifecycle stage because controls and evidence naturally attach to build, release, onboarding, runtime, and update checkpoints. Table 1 summarizes representative attacks and impacts. The taxonomy is intended to be operator-facing: each threat should map to an enforcement gate and an auditable evidence bundle.

*Table 1. Lifecycle-organized taxonomy of supply-chain threats for xApps/rApps.*

| Lifecycle stage | Representative attacks | Primary affected assets | Typical impact |
|---|---|---|---|
| Build / package | Compromised CI runner; poisoned dependency; malicious build script; build-cache poisoning | Source, build environment, dependencies, produced artifact | Integrity loss; hidden backdoors; reproducibility failure |
| Sign | Signing-key theft; unauthorized signing; signature stripping; forged attestations | Signing keys, certificates, provenance, release metadata | Bypass of verification; persistent trust compromise |
| Publish | Registry poisoning; tag overwrite; replay of vulnerable versions; metadata tampering | Registry, artifact index, versioning metadata | Distribution of malicious or outdated artifacts; rollback exposure |
| Onboard | Admission bypass; policy mis-binding; identity spoofing; weak evidence checks | Onboarding pipeline, app identity, runtime permissions | Privilege misassignment; malicious app installed; lateral-movement seed |
| Runtime | Privilege escalation; side-channel or data exfiltration; policy abuse; DoS of RIC services | Telemetry/control data, platform services, other apps | Confidentiality breach; availability degradation; unsafe control actions |
| Update / rollback | Tampered update; downgrade to vulnerable release; partial-rollout manipulation | Update channel, rollout policy, runtime state | Reintroduction of known vulnerabilities; integrity or availability incident |

In O-RAN, artifacts may include policy logic and AI/ML components, so integrity and provenance should cover more than container images alone. Because xApps can influence near-real-time control loops, availability and safety impacts matter alongside classic confidentiality and integrity concerns.

## 5 Zero-Trust Control Framework

The framework combines the lifecycle taxonomy, the NIST zero-trust policy and enforcement model, and WG11 security baselines into a mapping from threats to controls to verifiable evidence [2]-[6]. A key design choice is to treat evidence as first-class: controls that cannot be checked during onboarding or via continuous monitoring are weak for third-party app assurance.

### 5.1 Mapping zero-trust components to O-RAN enforcement points

In RIC deployments, natural policy enforcement points include API gateways, service meshes, O-Cloud admission controls, runtime isolation mechanisms, and the app manager that binds identities to permissions. Policy decision and administration roles typically sit in the SMO/Non-RT RIC domain where operator policy is authored and distributed to enforcement points [6].

### 5.2 WG11 transport, identity, and testing baseline

WG11 specifies security requirements and protocol profiles for applicable O-RAN components and interfaces, including secure communication, certificate-based authentication, and related conformance testing [2]-[4]. For RIC apps, the practical implication is that platform API access should be based on credential-bound identity and least-privilege authorization rather than network location alone. This paper reuses WG11-style validation artifacts and complements them with app supply-chain evidence such as SBOMs, signed metadata, and provenance attestations.

### 5.3 Threat-control-evidence mapping

Table 2 summarizes the proposed mapping. The last column is deliberately phrased as a baseline or complementary evidence anchor; the table should not be read as claiming that WG11 already standardizes every SBOM, SLSA, transparency-log, or anti-rollback requirement. Those items are proposed as app-centric evidence that can be layered onto WG11 conformance practice.

*Table 2. Condensed threat-control-evidence matrix.*

| **Threat** | **Control(s)** | **Verifiable evidence** | **WG11 baseline / complementary evidence anchor** |
|---|---|---|---|
| Build compromise | Harden CI; reviewed changes; isolated builds; provenance | Provenance attestation; build logs; CI identity | WG11 hardening and test evidence are complemented by build evidence [4], [8]-[10] |
| Dependency confusion | Pinned dependencies; allowlists; private registries; SBOM | SBOM; dependency-policy report; vulnerability scan | Dependency transparency complements interface and platform controls [2], [9] |
| Signing / registry compromise | Protected signing keys; immutable tags; signed metadata | Signature chain; registry audit log; optional transparency proof | Certificate and secure-transport baselines plus signing evidence [3], [13], [14] |
| Weak onboarding verification or excessive permissions | Verify signature, SBOM, and provenance; least privilege; approval workflow | Admission log; policy snapshot; RBAC/ABAC diff | Authorization, secured API, and audit-oriented tests [2]-[4] |
| Runtime lateral movement | Segmentation; mutual authentication where applicable; workload isolation; audit monitoring | Network policy; authentication configuration; workload policy; audit logs | Secured API and runtime security test practices [2]-[4] |
| Tampered update / downgrade | Signed releases; staged rollout; anti-rollback checks | Signed manifest; rollout log; rollback authorization | WG11 validation practice extended to update gates [4] |

## 6 RIC App Supply-Chain Assurance Profile

The RIC App Supply-Chain Assurance Profile is an operator-facing conformance rubric for procurement, onboarding, and audit. It bundles three evidence families: secure development practices, SBOM transparency, and build provenance/release integrity [8]-[10]. These evidence families complement, rather than replace, WG11 runtime and interface security baselines.

The levels below are not intended to redefine SLSA levels. Instead, they package SSDF, SBOM, and SLSA-style provenance evidence into a RIC-app assurance profile that is simple enough for admission policy. Operationally, a RIC app satisfies Level L only if each evidence family reaches Level L. The onboarding rule is therefore monotonic: allow the app only if its score is greater than or equal to the required level for the target environment.

*Table 3. RIC App Supply-Chain Assurance Profile.*

| **Level** | **SSDF evidence** | **SBOM / vulnerability transparency** | **Provenance / release integrity** |
|---|---|---|---|
| 0 | Ad hoc review only | No SBOM requirement | Unsigned or unverifiable artifacts |
| 1 | Documented SDLC policy; basic review | SBOM in SPDX or CycloneDX format; basic scan | Signed artifact; immutable version tags |
| 2 | Protected CI and release controls | SBOM plus allowlist and monitoring | Controlled build plus verifiable provenance, broadly aligned with SLSA Level 2 or higher |
| 3 | Separation of duties; hardened runners | SBOM plus VEX or documented exception workflow | Hardened build platform; anti-rollback controls; optional transparency logging |

Evidence can be shipped alongside the artifact as signed attestations, for example in-toto statements or other signed metadata produced by modern signing and transparency-backed tooling [13], [14]. The exact toolchain is intentionally not mandated; the operator should define trusted signers, expected builder identities, approved dependency sources, and required evidence levels according to deployment criticality.

## 7 Applicability Analysis and Prototype Checking Workflow

This section evaluates applicability rather than system performance. The goal is not to claim empirical deployment results, but to show how the threat-control-evidence mapping and assurance profile create auditable decision points. Three walkthroughs are used as negative test cases for a minimal checker.

## 7.1 Scenario walkthroughs

Scenario 1: tampered update of a third-party xApp. An attacker compromises a vendor release pipeline or distribution channel and publishes a tampered xApp image under a legitimate tag. Under the proposed profile, onboarding requires a valid signature bound to an authorized signer and provenance that matches the expected build identity and source inputs at Level 2 or above. The update is blocked unless the signature, digest, and provenance match policy.

Scenario 2: dependency confusion in an rApp build. A malicious package with the same name as an internal dependency is resolved into the build. Level 1 requires SBOM generation and a basic scan; Level 2 adds allowlist and monitoring requirements. The build is escalated or blocked when the SBOM or allowlist report reveals an unexpected external dependency.

Scenario 3: insider policy bypass leading to lateral movement. An insider with access to onboarding tools attempts to grant an xApp excessive permissions or disable segmentation. The proposed framework places this risk at policy decision and enforcement layers: onboarding binds app identity to a minimal role, policy changes require approval, and audit evidence is retained. The app is rejected or escalated if role, approval, or isolation evidence is missing.

*Table 4. Scenario-to-evidence mapping and recommended assurance levels.*

| Scenario | Key evidence checked | Recommended minimum level | Gate result |
|---|---|---|---|
| Tampered update | Signature and digest; provenance; admission log | Level 2 | Block unless signer and provenance match policy |
| Dependency confusion | SBOM; allowlist report; vulnerability scan | Level 1, with Level 2 preferred | Escalate or block on mismatch |
| Insider policy bypass | Policy diff; approvals; audit logs; isolation policy | Level 2-3 | Reject or escalate if approvals or isolation evidence are missing |

## 7.2 Minimal checker design

A minimal prototype can be implemented as a deterministic evidence checker. The input is a RIC app submission package containing an artifact digest, signature metadata, an SBOM, and provenance attestations. The verifier first checks completeness and authenticity. A policy checker then maps the results to the proposed assurance levels and produces one of three outcomes: Accept, Escalate, or Block.

*Figure 2. Minimal evidence-checking workflow: submission package -> evidence verifier -> policy checker -> assurance score -> Accept / Escalate / Block.*

*Table 5. Prototype checks and example gate logic.*

| Check item | Evidence | Rule example | Output |
|---|---|---|---|
| Artifact integrity | Signature and digest | Signer must be trusted and digest must match submitted manifest | Pass / fail |
| Dependency transparency | SBOM and policy report | Dependencies must match approved sources and declared exception policy | Pass / fail / escalate |
| Build provenance | Provenance attestation | Builder identity and source repository must match approved pipeline | Pass / fail |
| Runtime policy binding | Admission log and policy snapshot | Requested privileges must be no broader than approved profile | Pass / fail / escalate |
| Assurance gate | Combined results | Level 2 requires all core checks to pass | Accept / Escalate / Block |

# 8 Limitations and Future Empirical Metrics

The evaluation in this preprint is analytical. It does not measure operational overhead, false-accept or false-reject behavior, deployment latency, or inter-reviewer consistency across human assessors. Provenance verification also depends on operator-managed trust anchors, expected builder identities, and explicit policy definitions. These are implementation and governance choices rather than properties that can be inferred from the artifact alone.

Future work should implement the checker against a realistic RIC app package and report quantitative results. Table 6 lists concrete metrics that would directly address the current limitations.

*Table 6. Suggested empirical metrics for future validation.*

| Metric | What it measures | Why it matters |
|---|---|---|
| Verification latency | Time to check signatures, SBOMs, provenance, and policy bundle | Shows whether evidence checks fit operational onboarding cadence |
| Evidence completeness | Fraction of required artifacts present and machine-verifiable | Distinguishes incomplete submissions from failed security checks |
| Decision consistency | Agreement between policy-checker decisions and human reviewer decisions | Tests whether the rubric reduces subjective review variance |
| Detection coverage | Whether tampered updates, dependency confusion, and policy-bypass cases are blocked | Validates the negative test cases used in the paper |
| Operational overhead | Additional onboarding steps, manual review burden, and policy maintenance effort | Measures practicality for operators and platform providers |

## 9 Related Work

O-RAN security studies have examined risks introduced by open interfaces, cloud-native deployment, and programmability, while WG11 provides threat modeling and risk assessment guidance for the O-RAN ecosystem [5], [17], [18]. These works motivate the need for security controls around interfaces, components, and control-loop behavior.

Software supply-chain research and practice provide reusable primitives such as SSDF, SBOM standards, provenance frameworks, and end-to-end attestation systems [8]-[14]. These mechanisms are largely domain-agnostic. This paper connects them to RIC lifecycle gates, WG11-aligned baselines, and operator onboarding decisions. Unlike generic supply-chain frameworks, RIC apps may interact with RAN control loops and operator policy, so evidence must be both onboarding-verifiable and runtime-monitorable.

## 10 Conclusion

O-RAN's open application ecosystem makes xApps and rApps a plausible entry point for software supply-chain attacks. This paper proposed a lifecycle threat model, a WG11-aligned threat-control-evidence mapping, and an operator-facing assurance profile that bundles SSDF, SBOM, and SLSA-style provenance into incremental levels. The case-study walkthroughs show how these artifacts can support explicit onboarding decisions in zero-trust O-RAN deployments. The main next step is empirical validation through a working evidence checker, realistic RIC app packages, and quantitative measurements of verification cost and decision quality.